\documentclass[rnote,bibyear]{aa}  

\usepackage{graphicx}
\usepackage{txfonts}
\begin{document}

   \title{WISEA~J064750.85-154616.4: a new nearby L/T transition 
dwarf\thanks{based on observations with the Large Binocular Telescope (LBT)}}


\titlerunning{New L/T transition dwarf}

   \author{R.-D. Scholz
          \and
          G. Bihain
          \and
          J. Storm}

   \institute{Leibniz-Institut f\"ur Astrophysik Potsdam (AIP),
              An der Sternwarte 16, 14482 Potsdam, Germany\\
              \email{rdscholz@aip.de, gbihain@aip.de, jstorm@aip.de}
             }

   \date{Received 25 April 2014; accepted 4 June 2014 }

 
  \abstract
   {}
   {Our aim is to detect and classify previously overlooked brown dwarfs 
    in the solar neighbourhood.}
   {We performed a proper motion search among bright sources 
    observed with the Wide-field Infrared Survey Explorer (WISE)
    that are also seen in the Two Micron All Sky Survey (2MASS).
    Our candidates appear according to their red $J$$-$$K_s$ colours
    as nearby late-L dwarf candidates. 
    Low-resolution near-infrared (NIR) classification spectroscopy
    in the $HK$ band allowed us to get spectroscopic distance and 
    tangential velocity estimates.}
   {We have discovered a new L9.5 dwarf, WISEA~J064750.85-154616.4,
    at a spectroscopic distance of about 14~pc and with a tangential 
    velocity of about 11~km/s, typical of the Galactic thin disc population.
    We have confirmed another recently found L/T transition object 
    at about 10~pc, WISEA~J140533.13+835030.7, which we classified as L8 (NIR).}
   {}


   \maketitle

   \begin{figure}
   \centering
   \includegraphics[width=8.8cm]{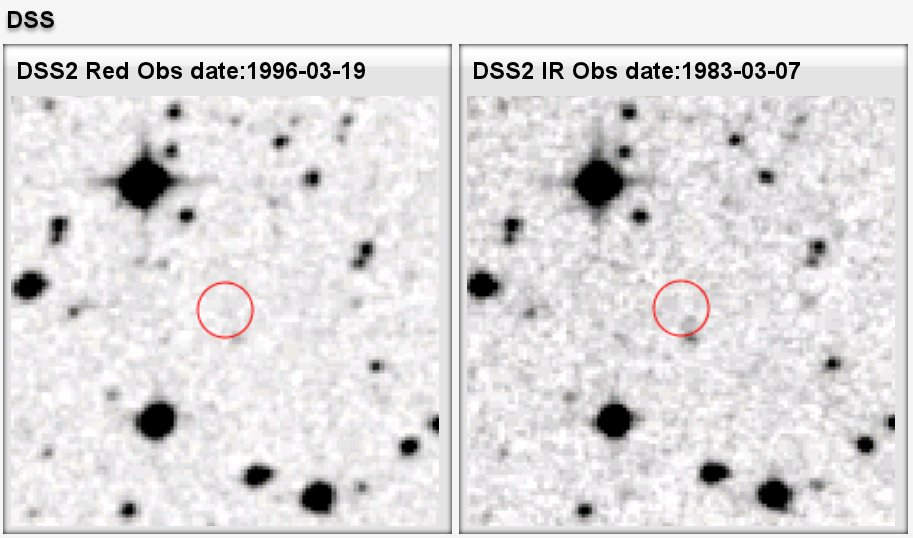}
   \includegraphics[width=8.8cm]{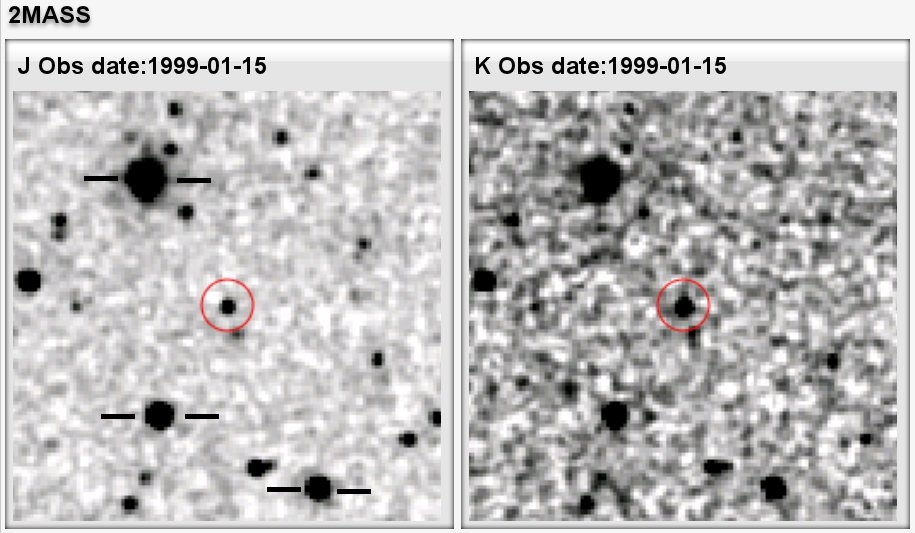}
   \includegraphics[width=8.8cm]{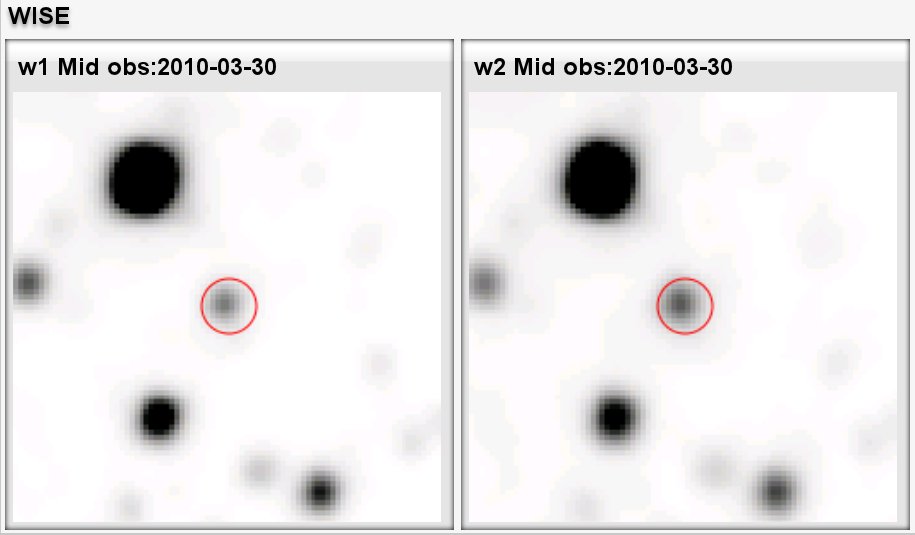}
      \caption{DSS $R$ and $I$, 2MASS $J$ and $K_s$, and WISE
               $w1$ and $w2$ finding charts (2$\times$2~arcmin$^2$,
               north is up, east to the left) for WISEA~J0647$-$1546.
               Circles are centered on target position at the 2MASS epoch.
               Three reference stars (see text) are marked in
               $J$ band.
              }
         \label{fig1_fc}
   \end{figure}


\section{Introduction}
\label{Sect_intro}

Proper motion searches continue to play an important role in the
search for still missing cool neighbours of the sun. 
For finding hidden brown dwarfs (BDs) in the solar neighbourhood by 
their high proper motion (HPM), it is essential
to use near-infrared (NIR) and mid-infrared (MIR) multi-epoch
observations. The NIR observations of the Two Micron All Sky Survey
(2MASS; Skrutskie et al.~\cite{skrutskie06}) with observing epochs 
between 1997 and 2001 have been
successfully combined with recent ($\approx$2010) MIR observations of the 
Wide-field Infrared Survey Explorer (WISE; 
Wright et al.~\cite{wright10}) to discover
nearby M, L and T dwarfs previously overlooked in colour-based
surveys (see e.g.
Gizis et al.~\cite{gizis11a}, \cite{gizis11b}, \cite{gizis12};
Castro \& Gizis~\cite{castro12};
Castro et al.~\cite{castro13};
Bihain et al.~\cite{bihain13};
Thompson et al.~\cite{thompson13};
Scholz~\cite{scholz14}).

The multiple epochs from WISE alone already allowed to find very large
numbers of previously unknown HPM objects (Luhman~\cite{luhman14a};
Kirkpatrick et al.~\cite{kirkpatrick14}). The majority
of these new HPM objects are according to their moderately red
colours M dwarfs in the extended solar neighbourhood.
But spectroscopic follow-up observations also revealed
many ultracool (L-type) subdwarfs (Wright et al.~\cite{wright14}; 
Kirkpatrick et al.~\cite{kirkpatrick14};
Luhman \& Sheppard~\cite{luhman14c}) among the new WISE HPM objects.
In addition, a few very close ($d$$<$10~pc) new L/T and M dwarf neighbours,
which were observed but not discovered before in 2MASS,
were found by Luhman~(\cite{luhman13}) 
and Kirkpatrick et al.~(\cite{kirkpatrick14}). The even cooler
Y-type BDs discovered thanks to WISE observations
(Cushing et al.~\cite{cushing11}; 
Kirkpatrick et al.~\cite{kirkpatrick12}, \cite{kirkpatrick13};
Cushing et al.~\cite{cushing14}) are not seen in 2MASS.
The recently discovered coolest BD (Luhman~\cite{luhman14b}),
ranging now as the third fastest among all HPM objects and 
fourth nearest among the solar neighbours, 
is undetected in the $J$ band down to 23rd magnitude.

In their HPM search for previously overlooked (in 2MASS) nearby BDs,
Scholz et al.~(\cite{scholz11}) and
Bihain et al.~(\cite{bihain13}) concentrated on relatively bright 
MIR ($w2$$\lesssim$13) BD candidates selected from the WISE 
preliminary and all-sky catalogues, respectively.
They looked for their 2MASS counterparts with significant proper motion
and colours typical of T dwarfs. Some of the candidates in the search
of Bihain et al.~(\cite{bihain13}) exhibited relatively red
NIR colours expected for late-L dwarfs. Here we report the results
of spectroscopic follow-up observations for three of these
candidates, one of which turned out to be a new L9.5 dwarf within about 
15~pc. For all objects described in this research note, we use their
AllWISE (Kirkpatrick et al.~\cite{kirkpatrick14}) designations.

   \begin{figure*}
   \centering
   \includegraphics[width=15.4cm]{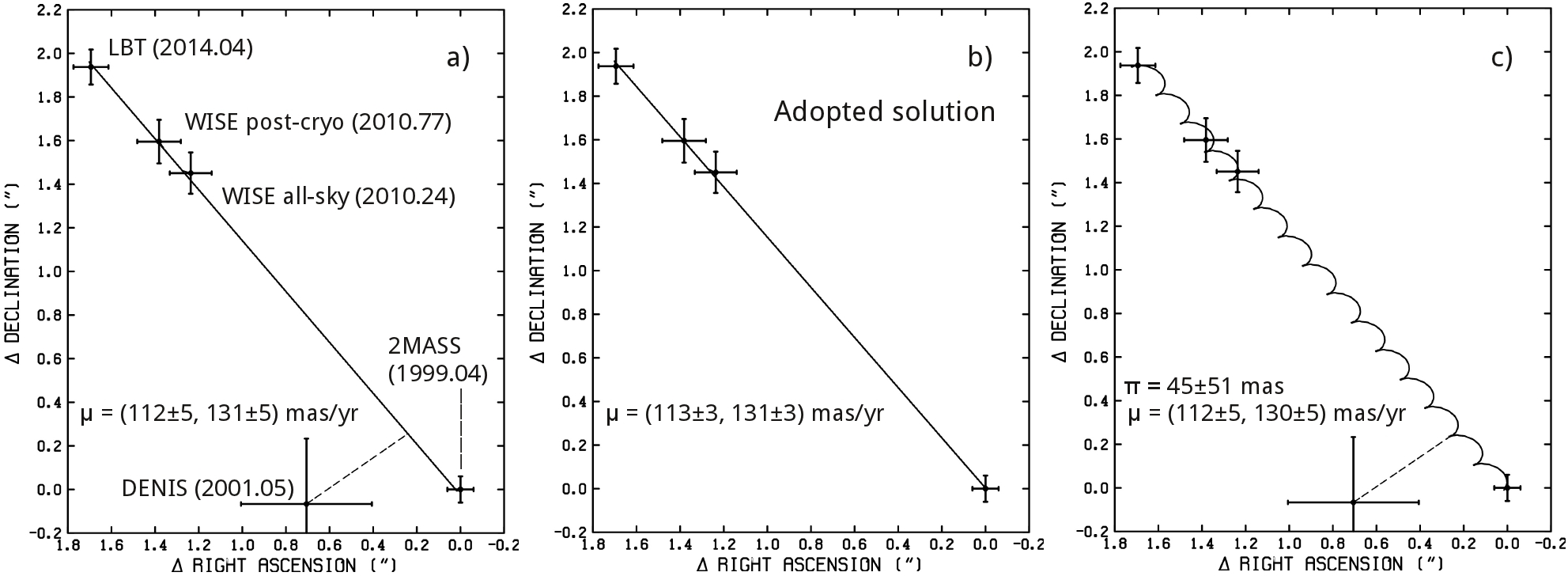}
      \caption{Proper motion and parallax solutions using the
               software of Gudehus~(\cite{gudehus01}): a) proper motion
               including the DENIS position with large error bars,
               b) proper motion from only four accurate positions
               (adopted solution), c) combined proper motion and
               parallax solution using all available positions.
              }
         \label{fig2_pmplx}
   \end{figure*}


\section{Candidate selection and proper motion}
\label{Sect_select}

The new object WISEA~J064750.85-154616.4 (hereafter WISEA~J0647$-$1546)
was found with the same selection criteria 
as described in Bihain et al.~(\cite{bihain13}). Its proper motion
was first obtained from the comparison of only two epochs in the 
WISE all-sky and 2MASS data. Later we further improved it by
including the positions measured in the WISE post-cryo data and 
in the Large Binocular Telescope (LBT) acquisition image 
observed for our spectroscopic follow-up 
(Sect.~\ref{Sect_nirspec}). Because of its red $J$$-$$K_s$ colour
(Table~\ref{table:1}), WISEA~J0647$-$1546 was considered as a
late-L dwarf candidate. Two more red candidates were selected 
for our spectroscopic follow-up (see Sect.~\ref{Sect_nirspec}).

Figure~\ref{fig1_fc} shows the field around the new nearby target as 
observed with the optical Digitized Sky Surveys (DSS), 
near-infrared (NIR) 2MASS and mid-infrared (MIR) WISE.
The object was not detected in the $I$-band measurements of the 
SuperCOSMOS Sky Surveys (SSS; Hambly et al.~\cite{hambly01}). But it can
be seen in the corresponding image (top right in Fig.~\ref{fig1_fc}) 
at the expected position at epoch 1983, close to
the south-west inner border of the circle. 
On the other hand, its proper motion leads to a
shifted position in north-east direction with respect to the circle 
centre at the later WISE epoch (bottom row).

The target also appears as a very faint object in the
DEep Near-Infrared Survey (DENIS; Epchtein et al.~\cite{epchtein97}).
However, the DENIS position deviates by about 0.8~arcsec from
the expected position at the DENIS observing epoch. The latter is
in addition not known exactly (see footnote in Table~\ref{table:1}).
We corrected the DENIS position of the target for the mean offset
of three reference stars close to the target (marked in $J$-band image 
of Fig.~\ref{fig1_fc}) with respect to their 2MASS positions, as we
also did for the positions measured in the LBT acquisition image.
These three stars have no significant proper motions ($<$10~mas/yr)
in the PPMXL (R\"oser, Demleitner \& Schilbach~\cite{roeser10}).

The corrected DENIS position (Table~\ref{table:1}) is still off by 
about 0.5~arcsec in right ascension and 
0.3~arcsec in declination from the expected position (see left
and right panels
in Fig.~\ref{fig2_pmplx}). Therefore we tried to measure the target
position in the DENIS $J$ and $K_s$ FITS images, but failed because 
the target image is almost completely buried in noise. We assigned
large error bars ($\pm$300~mas) to the (corrected) DENIS catalogue 
position, whereas the errors of the 2MASS, LBT, and two WISE positions
range between 60 and 100~mas. The linear proper motion fit with
and without the DENIS position does not change significantly, but
the proper motion error in the solution without DENIS is much 
smaller so that we adopted this as our final solution 
(Table~\ref{table:1}, Fig.~\ref{fig2_pmplx}, central panel). 
The four precise positions are consistent with a 
purely linear motion.
They are also consistent with a combined proper
motion and parallax solution, where the parallax and its error
are of the order of 50~mas (Fig.~\ref{fig2_pmplx} right panel).
However, we consider this parallax as unreliable not only due
to the large formal error but also because of
the small number of positions, their partly similar seasons, 
and the different NIR and MIR observations used.

The multi-epoch positions, the determined proper motion, and
the available NIR and MIR photometry of WISEA~J0647$-$1546 are listed
in Table~\ref{table:1}. The proper motion given in the AllWISE catalogue
is $\mu_{\alpha}\cos{\delta}$=$+$344$\pm$65, $\mu_{\delta}$=$+$312$\pm$66
mas/yr. Because of the very small time interval between the WISE observations
of about six months, it is much less precise and probably more affected
by parallactic motion than our proper motion.


\section{Near-infrared spectroscopic classification}
\label{Sect_nirspec}

Our target WISEA~J0647$-$1546 as well as two additional candidates
found in our survey were placed in a bad-weather backup programme 
at the LBT starting already in 2012. The two other objects are
WISEA~J140533.13+835030.7 (hereafter WISEA~J1405+8350), which has been
meanwhile discovered and described as an L9 dwarf (NIR) at a distance 
of 9.7$\pm$1.7~pc by Castro et al.~(\cite{castro13}), and
WISEA~J042144.33$+$192943.8 (hereafter WISEA~J0421$+$1929).
For these relatively red ($J$$-$$K_s$$>$1.5)
targets with $K_s$$<$14, we used the NIR spectrograph LUCI~1
(Mandel et al.~\cite{mandel08}; Seifert et al.~\cite{seifert10};
Ageorges et al.~\cite{ageorges10}) in long-slit
spectroscopic mode only with the $HK$ (200 lines/mm + order separation 
filter) grating.
The three objects, WISEA~J0647$-$1546, WISEA~J1405+8350, and 
WISEA~J0421$+$1929, were observed on 2014-Jan-13, 2014-Jan-12,
and 2014-Jan-15 (UT) with total integration times of 12, 8, and 12 minutes, 
respectively. Acquisition images were taken with the $K_s$
filter with exposure times of 15, 5, and 15 seconds, respectively.
The seeing was $\approx$2~arcsec and the slit width 
was 1~arcsec. The spectroscopic observations and data reduction were
otherwise as described by Bihain et al.~(\cite{bihain13})
with the difference that here the frames were illumination corrected 
before sky subtraction and the wavelength calibration was obtained using 
the sky emission OH lines, with an accuracy of about 0.9\AA{}.

   \begin{figure*}
   \sidecaption
   \includegraphics[width=12.1cm]{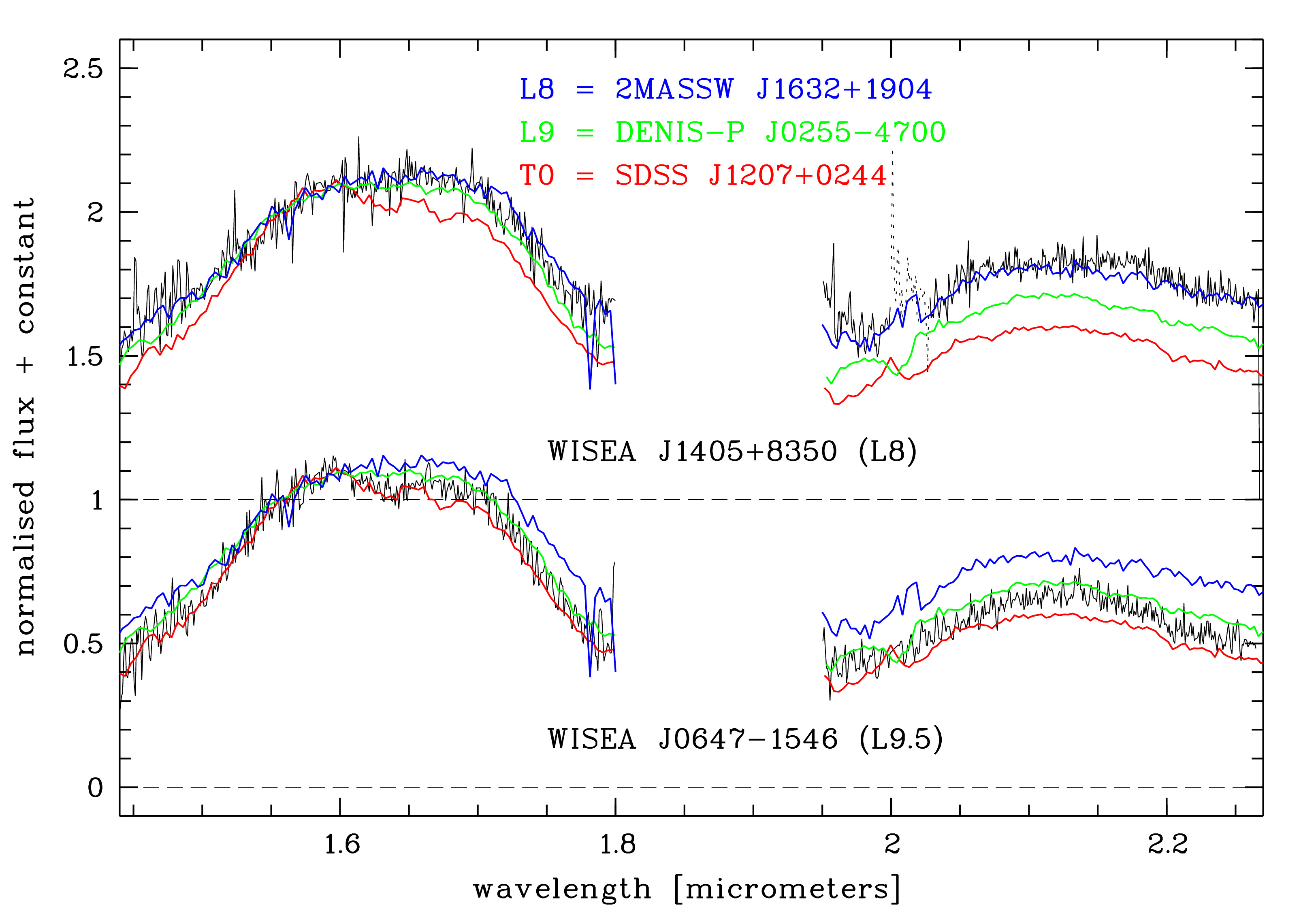}
      \caption{LBT/LUCI spectra (black)
of WISEA~J0647$-$1546 (bottom) and WISEA~J1405+8350 (top)
overplotted with lower resolution NIR standard spectra:
2MASSW~J1632$+$1904 (L8, blue; Burgasser~\cite{burgasser07}),
DENIS-P~J0255$-$4700 (L9, green; Burgasser et al.~\cite{burgasser06}), and
SDSS~J1207$+$0244 (T0, red; Looper, Kirkpatrick, and Burgasser~\cite{looper07}).
The spectrum of WISEA~J1405+8350 is affected by strong residuals of telluric 
        absorption correction in the region of 2.00..2.03 micrometers (dotted
        line), as the A0V standard star used for the correction
        was observed at higher airmass than the target.
              }
         \label{fig3_spec}
   \end{figure*}

In Fig.~\ref{fig3_spec}, we show the LBT/LUCI spectra of the
new object WISEA~J0647$-$1546 and of WISEA~J1405+8350, together
with L8, L9, and T0 standard spectra.  
For WISEA~J0647$-$1546, the $H$-band spectrum fits slightly better
with the T0 than with the L9 standard spectrum, whereas in the 
$K$ band it lies right in between the L9 and T0 standard spectra.
All three measured spectral indices (Table~\ref{table:1})
are consistent with a spectral type of T0 as shown by 
Burgasser et al.~(\cite{burgasser06}, his Table 5).
The spectral index/spectral type relation for the H$_2$O-$H$
index provided by Burgasser~(\cite{burgasser07}, their Table 3) 
leads to a spectral type of L9.3, the relation for the CH$_4$-$H$ 
index is only applicable in the T1-T8 range, and that for
the CH$_4$-$K$ index yields an L9.5 type. Due to the
good agreement between the results from direct comparison 
with standard spectra and from the spectral indices, we
classified WISEA~J0647$-$1546 as an L9.5 dwarf with an uncertainty
of half a spectral subclass.

Dupuy \& Liu~(\cite{dupuy12}) provided mean 2MASS and WISE
absolute magnitudes for M, L and T spectral types. Because
there are no data for L9.5 dwarfs in their Tables 16 and 18,
we used the average absolute $JHK_s$ magnitudes of L9 and T0 dwarfs
and the absolute $w1w2$ magnitudes of L9 dwarfs to compute
spectroscopic distances of WISEA~J0647$-$1546. They range from 
13.1~pc (from $w1$) to 14.8~pc (from $H$) with a mean distance
of 13.9$^{+3.6}_{-2.9}$~pc, where we conservatively assumed
a large uncertainty of 0.5~mag in the absolute magnitude. The 
resulting tangential velocity is about 11~km/s.

The spectrum of WISEA~J1405+8350
fits best with the L8 standard (slightly closer to L8 than L9 in $H$ band
and well-fitted by L8 in $K$ band), whereas Castro et al.~(\cite{castro13})
classified it as L8 in the optical but L9 in the NIR. From three
spectral indices, which we were able to measure for WISEA~J1405+8350
(H$_2$O-$H$$=$0.709, CH$_4$-$H$$=$1.051, CH$_4$-$K$$=$0.874),
the first leads to L6.2, the second cannot be applied,
and the third corresponds to a spectral type of L8.1
according to Burgasser~(\cite{burgasser07}). Based
on the direct comparison with the standard spectra and the
results from spectral indices, we classified WISEA~J1405+8350 as
an L8 dwarf.

The spectrum of
WISEA~J0421$+$1929 (not shown) represents a noisy red continuum.
From the location of WISEA~J0421$+$1929 close to the famous star
T~Tauri and overlapping with a large molecular cloud, we conclude
that this is a strongly reddened early-type star in the background.
We measured a proper motion of ($\mu_{\alpha}\cos{\delta}$, $\mu_{\delta}$) =
($+$56$\pm$12, $-$110$\pm$11)~mas/yr. This translates to a
tangential velocity of more than about 100~km/s, if we assume
WISEA~J0421$+$1929 to move behind the nebula and T~Tauri, which
lies at a distance of about 180~pc as known from Hipparcos
(van Leeuwen~\cite{vanleeuwen07}). 

%
\begin{table}
\caption{Parameters of WISEA~J0647$-$1546} 
\label{table:1}      
\centering                          
\begin{tabular}{lc}        
\hline\hline                 
Parameter   & WISEA~J0647$-$1546   \\ 
\hline                        
2MASS $\alpha$ (J2000)          &     06 47 50.763 \\
2MASS $\delta$ (J2000)          &  $-$15 46 18.00  \\
2MASS epoch                     &        1999.041  \\
DENIS $\alpha$ (J2000)\tablefootmark{a} & 06 47 50.812 \\
DENIS $\delta$ (J2000)\tablefootmark{a} & $-$15 46 18.06\\
DENIS epoch\tablefootmark{b}    &        2001.047  \\
WISE all-sky $\alpha$ (J2000)   &     06 47 50.848 \\
WISE all-sky $\delta$ (J2000)   &  $-$15 46 16.55  \\
WISE all-sky epoch              &        2010.244  \\
WISE post-cryo $\alpha$ (J2000)\tablefootmark{c} &     06 47 50.858 \\
WISE post-cryo $\delta$ (J2000)\tablefootmark{c} &  $-$15 46 16.40  \\
WISE post-cryo epoch            &        2010.770  \\
LBT $\alpha$ (J2000)            &     06 47 50.880 \\
LBT $\delta$ (J2000)            &  $-$15 46 16.06  \\
LBT epoch                       &        2014.036  \\
$\mu_{\alpha}\cos{\delta}$ (mas/yr) & $+$113$\pm$3 \\
$\mu_{\delta}$ (mas/yr)             & $+$131$\pm$3 \\
\hline
2MASS $J$ (mag)                 & 15.31$\pm$0.05 \\
2MASS $H$ (mag)                 & 14.29$\pm$0.05 \\
2MASS $K_s$ (mag)               & 13.74$\pm$0.06 \\
DENIS $J$ (mag)                 & 16.04$\pm$0.21 \\
DENIS $K_s$ (mag)               & 13.90$\pm$0.25 \\
AllWISE $w1$ (mag)              & 13.04$\pm$0.02 \\
AllWISE $w2$ (mag)              & 12.52$\pm$0.02 \\
AllWISE $w3$ (mag)              & 11.59$\pm$0.22 \\
\hline
H$_2$O-$H$ (SpT)\tablefootmark{d}                & 0.622 (T0) \\
CH$_4$-$H$ (SpT)\tablefootmark{d}                & 0.986 (T0) \\
CH$_4$-$K$ (SpT)\tablefootmark{d}                & 0.800 (T0) \\
Spectral type                   & L9.5$\pm$0.5 \\
$d_{spec}$ (pc)                 & 13.9$^{+3.6}_{-2.9}$ \\
$v_{tan}$ (km/s)                & 11$^{+3}_{-2}$ \\
\hline                                   
\end{tabular}
\tablefoot{
\tablefoottext{a}{The original DENIS coordinates of the target were
corrected for the mean offsets of three reference stars
around the target with respect to their 2MASS coordinates.
The DENIS data were not used for the final proper motion
determination.
}
\tablefoottext{b}{The DENIS catalogue gives an epoch of 2000.47,
whereas according to the DENIS FITS images it is 2001.47.}
\tablefoottext{c}{Mean position from 15 single exposures.}
\tablefoottext{d}{Spectral index (and corresponding spectral type) 
as defined in Burgasser et al.~(\cite{burgasser06}).}
}
\end{table}


\section{Conclusions and discussion}
\label{Sect_concl}

Best et al.~(\cite{best13}) have shown that previous surveys for
L/T transition dwarfs have been incomplete. We were able to 
discover a new L/T transition dwarf within about 15~pc
from the sun. With its spectral type of L9.5 that we found from
LBT/LUCI $HK$-band spectroscopy, WISEA~J0647$-$1546 is a promising
target for trigonometric parallax programmes, as for this type
except for 2MASSI~J0328426$+$230205 (= WISEA~J032842.65$+$230204.5
with a preliminary parallax of
33.1$\pm$4.1~mas measured by Vrba et al.~\cite{vrba04})
no reference values are available so far (Dupuy \& Liu~\cite{dupuy12}). 
Our new L9.5 dwarf WISEA~J0647$-$1546 is brighter than all six
L9.5 dwarfs (three of them with very uncertain types) listed in the 
DwarfArchives (Gelino, Kirkpatrick \& Burgasser~\cite{gelino12}).
With $K_s$$\approx$13.7, it is as bright as PSO~J140.2308+45.6487 
(= WISEA~J092055.41$+$453855.9) newly classified by 
Best et al.~(\cite{best13}) as L9.5 dwarf (with signs of spectral 
variability), which was 
first photometrically estimated as mid-L dwarf by
Aberasturi et al.~(\cite{aberasturi11}) and then
classified as an L9 and weak binary candidate by
Mace et al.~(\cite{mace13}).
Although from our spectrum there are no hints on a possible close 
companion, WISEA~J0647$-$1546 may also be worth high-resolution imaging
observations and variability analysis 
(cf. Biller et al.~\cite{biller13}; 
Burgasser et al.~\cite{burgasser14}; Gelino et al.~\cite{gelino14}).

As one of the rare bright L/T transition objects, the new 
L9.5 dwarf WISEA~J0647$-$1546 could be important 
for the study of cloud evolution of L and T dwarfs.
Therefore, it is again an interesting target
for time-resolved photometry (see e.g. Radigan et al.~\cite{radigan14};
Wilson, Rajan \& Patience~\cite{wilson14})
and spectroscopy at high signal-to-noise ratio as carried out e.g. by
Apai et al.~(\cite{apai13}) and Buenzli et al.~(\cite{buenzli14}). 
Potentially, it could even be included in future Doppler imaging
studies such as proposed by Crossfield et al.~(\cite{crossfield14}). 
Neither the spectrum nor the colours of WISEA~J0647$-$1546 show 
signs of youth (triangular peak in the $H$ band and very 
red $J$$-$$K_s$$\gtrsim$2.3 as observed e.g. by 
Schneider et al.~\cite{schneider14}), 
and its proper motion and distance are not consistent with membership
in one of the known young nearby associations 
(cf. Gagn{\'e} et al.~\cite{gagne14}).
The colours of WISEA~J0647$-$1546 are typical of its
spectral type and do not hint at other peculiarities,
as can be seen in comparison with the L/T
transition objects listed in Dupuy \& Liu~(\cite{dupuy12}).

For WISEA~J1405+8350, Castro et al.~(\cite{castro13}) obtained
an optical spectral type of L8 and a NIR type of L9 from high
signal-to-noise
$J$-band spectroscopy. In their Fig.~5, they also showed an 
$HK$-band spectrum, which they described as being consistent 
with the L9 type. They used the same L8 and L9 standards for
comparison as we did in Fig.~\ref{fig3_spec}. We find their
$HK$-band spectrum fits in between the L8 and L9 
standards, whereas ours fits better with the L8 standard as do our 
measured spectral indices. Nevertheless, our bad-weather
spectrum of WISEA~J1405+8350 is as noisy as their $HK$-band 
spectrum so that we think their more accurate classification 
in the $J$ band is more reliable.

For the reddened star WISEA~J0421$+$1929, the estimated velocity 
is much higher than the typical velocity of runaway stars and 
does not point in a direction out of the molecular cloud. 
Therefore a location behind but not in the cloud and a 
Galactic thick disc or halo membership seems to be plausible.
This conclusion is in agreement with recent findings of 
Esplin, Luhman \& Mamajek~(\cite{esplin14}), who have
spectroscopically classified WISEA~J0421$+$1929 as a non-member 
of Taurus with a spectral type of $<$M0. 


\begin{acknowledgements}
The authors thank Jochen Heidt, Barry Rothberg, and the observers at 
the LBT, D. Rosario and E. Wuyts, for assistance during the preparation 
and execution of LUCI observations, Adam Burgasser for providing 
template spectra at http://pono.ucsd.edu/$\sim$adam/browndwarfs/spexprism
and the referee, {\'E}tienne Artigau, for helpful hints, mainly concerning the
discussion of our results. 
This research has made use of the
NASA/IPAC Infrared Science Archive, which is operated by the Jet Propulsion
Laboratory, California Institute of Technology, under contract with the
National Aeronautics and Space Administration,
of data products from WISE,
which is a joint project of the University of California,
Los Angeles, and the Jet Propulsion Laboratory/California Institute of
Technology, funded by the National Aeronautics and Space Administration,
and from 2MASS.
We have also extensively used the M, L, T, and Y dwarf compendium housed at
DwarfArchives.org, the SIMBAD database and the VizieR catalogue
access tool operated at the CDS/Strasbourg,
including the CDS xMatch service.  
\end{acknowledgements}



\begin{thebibliography}{}

\bibitem[2011]{aberasturi11}
Aberasturi, M., Solano, E., \& Mart{\'{\i}}n, E.~L.\ 2011, A\&A, 534, L7

\bibitem[2010]{ageorges10}
Ageorges, N., Seifert, W., J\"utte, M., et al.\ 2010, SPIE, 7735, 53

\bibitem[2013]{apai13}
Apai, D., Radigan, J., Buenzli, E., et al.\ 2013, ApJ, 768, 121

\bibitem[2013]{best13}
Best, W.~M.~J., Liu, M.~C., Magnier, E.~A., et al.\ 2013, ApJ, 777, 84

\bibitem[2013]{bihain13}
Bihain, G., Scholz, R.-D., Storm, J., \& Schnurr, O.\ 2013, A\&A, 557, A43

\bibitem[2013]{biller13}
Biller, B.~A., Crossfield, I.~J.~M., Mancini, L., et al.\ 2013, ApJ, 778, L10

\bibitem[2014]{buenzli14}
Buenzli, E., Apai, D., Radigan, J., Reid, I.~N., \& Flateau, D.\ 2014, ApJ, 782, 77

\bibitem[2006]{burgasser06}
Burgasser, A.~J.,
Geballe, T.~R., Leggett, S.~K., Kirkpatrick, J.~D.,
\& Golimowski, D.~A.\ 2006, ApJ, 637, 1067

\bibitem[2007]{burgasser07}
Burgasser, A.~J.\ 2007, ApJ, 659, 655

\bibitem[2014]{burgasser14}
Burgasser, A.~J., Gillon, M., Faherty, J.~K., et al.\ 2014, ApJ, 785, 48

\bibitem[2012]{castro12}
Castro, P.~J., \& Gizis, J.~E.\ 2012, ApJ, 746, 3

\bibitem[2013]{castro13}
Castro, P.~J., Gizis, J.~E., Harris, H.~C., et al.\ 2013, ApJ, 776, 126

\bibitem[2014]{crossfield14}
Crossfield, I.~J.~M.\ 2014, arXiv:1404.7853

\bibitem[2011]{cushing11}
Cushing, M.~C., Kirkpatrick, J.~D., Gelino, C.~R., et al.\ 2011, ApJ, 743, 50

\bibitem[2014]{cushing14}
Cushing, M.~C.,
Kirkpatrick, J.~D., Gelino, C.~R., et al.\ 2014, AJ, 147, 113

\bibitem[2012]{dupuy12}
Dupuy, T.~J., \& Liu, M.~C.\ 2012, ApJ Suppl. Ser., 201, 19

\bibitem[1997]{epchtein97}
Epchtein, N., de Batz, B., Capoani, L., et al.\ 1997, The Messenger, 87, 27

\bibitem[2014]{esplin14}
Esplin, T.~L., Luhman,
K.~L., \& Mamajek, E.~E.\ 2014, ApJ, 784, 126

\bibitem[2014]{gagne14}
Gagn{\'e}, J., Lafreni{\`e}re, D., Doyon, R., Malo, L.,
\& Artigau, {\'E}.\ 2014, ApJ, 783, 121

\bibitem[2012]{gelino12}
Gelino, C.~R., Kirkpatrick, J.~D., \& Burgasser, A.~J.\ 2012,
online database for 1281 L, T, and Y dwarfs at DwarfArchives.org
(Last update 6 November 2012)

\bibitem[2014]{gelino14}
Gelino, C.~R., Smart, R.~L., Marocco, F., et al.\ 2014, AJ, 148, 6

\bibitem[2011a]{gizis11a}
Gizis, J.~E., Burgasser,
A.~J., Faherty, J.~K., Castro, P.~J., \& Shara, M.~M.\ 2011a, AJ, 142, 171

\bibitem[2011b]{gizis11b}
Gizis, J.~E., Troup, N.~W., \& Burgasser, A.~J.\ 2011b, ApJ, 736, L34

\bibitem[2012]{gizis12}
Gizis, J.~E., Faherty, J.~K., Liu, M.~C., et al.\ 2012, AJ, 144, 94

\bibitem[2001]{gudehus01}
Gudehus, D.~H.\ 2001, Bulletin of the American Astronomical Society, 33, 850

\bibitem[2001]{hambly01}
Hambly, N.~C.,  MacGillivray, H.~T., Read M.~A., et al.\ 2001, MNRAS, 326, 1279

\bibitem[2012]{kirkpatrick12}
Kirkpatrick, J.~D., Gelino, C.~R., Cushing, M.~C., et al.\ 2012, ApJ, 753, 156

\bibitem[2013]{kirkpatrick13}
Kirkpatrick, J.~D.,
Cushing, M.~C., Gelino, C.~R., et al.\ 2013, ApJ, 776, 128

\bibitem[2014]{kirkpatrick14}
Kirkpatrick, J.~D., Schneider, A., Fajardo-Acosta, S., et al.\ 2014, ApJ, 783, 122

\bibitem[2007]{looper07}
Looper, D.~L., Kirkpatrick, J.~D., \& Burgasser, A.~J.\ 2007, AJ, 134, 1162

\bibitem[2013]{luhman13}
Luhman, K.~L.\ 2013, ApJ, 767, L1

\bibitem[2014a]{luhman14a}
Luhman, K.~L.\ 2014a, ApJ, 781, 4

\bibitem[2014b]{luhman14b}
Luhman, K.~L.\ 2014b, ApJ, 786, L18

\bibitem[2014]{luhman14c}
Luhman, K.~L., \& Sheppard, S.\ 2014, ApJ, 787, 126

\bibitem[2013]{mace13}
Mace, G.~N., Kirkpatrick, J.~D., Cushing, M.~C., et al.\ 2013,
ApJ Suppl. Ser, 205, 6

\bibitem[2008]{mandel08}
Mandel, H., Seifert, W., Hofmann, R., et al.\ 2008, SPIE, 7014, 124

\bibitem[2014]{radigan14}
Radigan, J., Lafreni{\`e}re, D., Jayawardhana, R., \& Artigau, E.\ 2014, arXiv:1404.3247

\bibitem[2010]{roeser10}
R\"oser, S., Demleitner, M., Schilbach, E.\, 2010, AJ, 139, 2440

\bibitem[2014]{schneider14}
Schneider, A.~C., Cushing, M.~C., Kirkpatrick, J.~D., et al.\ 2014, AJ, 147, 34

\bibitem[2011]{scholz11}
Scholz, R.-D., Bihain, G., Schnurr, O., \& Storm, J.\ 2011, A\&A, 532, L5

\bibitem[2014]{scholz14}
Scholz, R.-D.\ 2014, A\&A, 561, A113

\bibitem[2010]{seifert10}
Seifert, W., Ageorges, N., Lehmitz, M., et al.\ 2010, SPIE, 7735, 256

\bibitem[2006]{skrutskie06}
Skrutskie, M.~F., Cutri, R.~M., Stiening, R., et al.\ 2006, AJ, 131, 1163

\bibitem[2013]{thompson13}
Thompson, M.~A.,
Kirkpatrick, J.~D., Mace, G.~N., et al.\ 2013, PASP, 125, 809

\bibitem[2007]{vanleeuwen07}
van Leeuwen, F.\ 2007, A\&A, 474, 653

\bibitem[2004]{vrba04}
Vrba, F.~J., Henden, A.~A., Luginbuhl, C.~B., et al.\ 2004, AJ, 127, 2948

\bibitem[2014]{wilson14}
Wilson, P.~A., Rajan, A., \& Patience, J.\ 2014, arXiv:1404.4633

\bibitem[2010]{wright10}
Wright, E.~L., Eisenhardt, P.~R.~M., Mainzer, A.~K., et al.\ 2010, AJ, 140, 1868

\bibitem[2014]{wright14}
Wright, E.~L., Kirkpatrick, J.~D., Gelino, C.~R., et al.\ 2014, AJ, 147, 61

\end{thebibliography}
\end{document}